\documentclass{siamart220329}

\usepackage{amsmath}
\usepackage{amssymb}

\usepackage{graphicx}
\usepackage{dcolumn}
\usepackage{bm}

\usepackage{hyperref}
\usepackage{subcaption}
\usepackage{siunitx}
\usepackage{mhchem}

\usepackage{xcolor}

\begin{document}


\title{Learning inducing points and uncertainty on molecular data by scalable variational Gaussian processes}


\author{
Mikhail Tsitsvero 
\thanks{Institute for Chemical Reaction Design and Discovery (WPI-ICReDD), Hokkaido University,
Sapporo 001-0021, Japan, \\ 
email: \email{tsitsvero@gmail.com}}
%
\and 
Mingoo Jin 
\thanks{Institute for Chemical Reaction Design and Discovery (WPI-ICReDD), Hokkaido University,
Sapporo 001-0021, Japan}
\and
Andrey Lyalin 
\thanks{Department of Chemistry, Faculty of Science, Hokkaido University, Sapporo 060-0810,
Japan; Research Center for Energy and Environmental Materials (GREEN), National Institute for Materials Science, Namiki 1-1, Tsukuba 305-0044, Japan. }}

\maketitle

\begin{abstract}
  Uncertainty control and scalability to large datasets are the two main issues for the deployment of Gaussian process (GP) models within the autonomous machine learning-based prediction pipelines in material science and chemistry. One way to address both of these issues is by introducing the latent inducing point variables and choosing the right approximation for the marginal log-likelihood objective. Here, we empirically show that variational learning of the inducing points in a molecular descriptor space improves the prediction of energies and atomic forces on two molecular dynamics datasets. First, we show that variational GPs can learn to represent the configurations of the molecules of different types that were not present within the initialization set of configurations. We provide a comparison of alternative log-likelihood training objectives and variational distributions. Among several evaluated approximate marginal log-likelihood objectives, we show that predictive log-likelihood provides excellent uncertainty estimates at the slight expense of predictive quality. Furthermore, we extend our study to a large molecular crystal system, showing that variational GP models perform well for predicting atomic forces by efficiently learning a sparse representation of the dataset.
\end{abstract}


\section{\label{sec:intro} Introduction}
Machine learning methods can significantly speed up the computational prediction of the physical and chemical properties of atomic systems by substituting the time-consuming electronic structure calculations with a surrogate predictive model. As predictive models for physical properties, such as energies and force fields, tend to become the active-learning components of the larger pipelines for material and chemical reaction design \cite{sbailo2022nomad,simm2018exploration, unsleber2022chemoton, maeda2016artificial,ceriotti2022beyond}, it is essential for such models to autonomously reason about the uncertainty of predictions. Although significant progress has been made in designing accurate models for approximating the potential energy surfaces \cite{bartok2010gaussian, vandermause2021active, chmiela2017machine, chmiela2018towards, chmiela2019sgdml, schutt2020machine}, robust uncertainty control is still in an under-developed state to be suitable for direct use in the systems with autonomous reasoning. Uncertainty quantification is typically addressed either by Bayesian model averaging, see e.g. \cite{wen2020uncertainty, imbalzano2021uncertainty} or by directly fitting the uncertainty quantification parameters of the predictive distribution, for example, within the framework of Gaussian processes \cite{musil2019fast, simm2018error, bartok2022improved,tsitsvero2023nmr}. In this work, we follow the second route and address the problem of learning the uncertainty parameters by scalable Gaussian processes. For an extensive review and applications of Gaussian processes to materials and molecules, we refer the reader to Ref. \cite{deringer2021gaussian}. 

Gaussian processes (GP) are often deemed the gold standard in machine learning, however, their application is hindered by poor scalability and over-estimation of a true uncertainty \cite{jankowiak2020parametric}. 
Parameters of the GPs are typically learned by maximizing the marginal log-likelihood objective function by the backpropagation method \cite{griewank2008evaluating} and stochastic gradient descent \cite{hoffman2013stochastic}, \cite{robbins1951stochastic}. Evaluation of exact marginal log-likelihood objective is computationally expensive for large datasets, therefore, several scalable approximations of the marginal log-likelihood were proposed recently \cite{knoblauch2019robust, jankowiak2020parametric} that are based on the Evidence Lower BOund (ELBO) inequalities. One way to address the scalability issue of the exact GPs is to sparsify them by introducing latent inducing point variables \cite{titsias2009variational}. In the following, we refer to the GP model augmented by the variational parameters that are learned by the stochastic gradient descent optimization of some objective function as the Stochastic Variational Gaussian Process (SVGP) model.  

The main contribution of our work is the application of SVGP framework to fit the potential energies and atomic forces of molecular systems with continuous molecular descriptors \cite{bartok2013representing}. We start by describing how a GP model can be augmented with inducing point latent variables to achieve scalability and improved uncertainty control. Then we present several options for approximating the marginal log-likelihood objective as well as outline options for modeling the variational distributions over the inducing variables. We illustrate the performance of the variational GP models with two numerical examples. First, we fit the potential energies of molecular configurations from Molecular Dynamics (MD) trajectories of three structural isomers of a \num{12}-atom molecule. We evaluate a collection of scalable GP models in terms of both the prediction quality on the test sets and the ability to separate the uncertainty for the training and test configurations. We demonstrate that learning the inducing points significantly improves prediction quality and show that inducing points initialized on the conformations of one kind of molecule can learn to represent the conformations of different kinds of molecules. Second, we illustrate that the SVGP framework also performs well for fitting the atomic forces. We provide an example of fitting the atomic forces of a large \num{528}-atom molecular crystal \cite{doi:10.1021/jacs.3c08909} by the SVGP framework and show that predictions by the sparsified variational models are on par with the predictions by the exact GP models.


Additionally, in the last section, we briefly discuss whether a GP model makes predictions on test configurations in high-dimensional descriptor space by interpolating within the training set of data points. By following the arguments by \cite{balestriero2021learning} we show, similarly to \cite{zeni2022exploring}, that most test samples indeed fall outside of the convex hull of densely sampled training points in the high-dimensional descriptor space, and therefore conclude that the GP model almost always makes predictions in the extrapolation regime.

\section{Fitting molecular data with scalable Gaussian processes}
\subsection{Invariant Descriptors}

To construct the rotation and translation invariant representations of atomic configurations in 3D space various methods were proposed where the data representation is either learned jointly with model parameters or constructed deterministically. Here we employ the widely used Smooth Overlap of Atomic Positions (SOAP) continuous and differentiable descriptors \cite{bartok2013representing, de2016comparing}. We briefly descirbe the construction of the invariant vectors. The atomic structure $\mathcal{A}$, composed of atoms with atomic numbers $Z$, is described by the set of atomic positions $\{ \mathbf{R}_i \}_{i \in \mathcal{A}}$ with respect to some selected origin which is referred to as the center. To build a representation of the entire atomic structure, we first select a collection of centers, capture the local environments around each center and combine them into global representation, see e.g. Ref. \cite{ceriotti2020machine} for details. The local atomic structure can be mapped to smooth atomic density by
$\rho^{Z}(\mathbf{r})=\sum_{i}^{\left|Z_{i}\right|} e^{-1 / 2 \sigma^{2}\left|\mathbf{r}-\mathbf{R}_{i}\right|^{2}} f_{cut}(\mathbf{r})$, where $f_{cut}$ is a cutoff function allowing to take into account only the atoms within short range $[ 0, r_{cut} ]$ from the center. To build the simplest invariant representation vector, the atomic density is first projected onto the set of radial and spherical basis functions by
\begin{equation}
  c_{n l m}^{Z}=\int_{\mathbb{R}^{3}} g_{n}(r) Y_{l m}(\theta, \phi) \rho^{Z}(\mathbf{r}) \mathrm{~d} V.
\end{equation}
Then, the simplest symmetry-invariant representation for a chosen center, referred to as SOAP power spectrum, is given by summation over the magnetic quantum number indices $m$:
\begin{equation}
  p_{n n^{\prime} l}^{Z_{1} Z_{2}}=\pi \sqrt{\frac{8}{2 l+1}} \sum_{m} \left(c_{n l m}^{Z_{1}}\right)^{*}\left(c_{n^{\prime} l m}^{Z_{2}}\right).
\end{equation}
As a radial basis set one can choose, for example, the Gaussian Type Orbital (GTO) basis functions. Depending on the complexity of the molecular dataset and size of the atomic structures, only the expansion coefficients with radial and spherical indices of the orders up to $n_{max}$ and $l_{max}$ are considered. A simple way to build a vector representing the entire structure is just to average the local power spectra over all the centers indexed by $i$
\begin{equation}
  \mathbf{x} \propto \sum_{i} \sum_{m}\left(c_{n l m}^{i, Z_{1}}\right)^{*}\left(c_{n^{\prime} l m}^{i, Z_{2}}\right).
  \label{eq:outer sum}
\end{equation}
Since all maps required to produce the feature vector $\mathbf{x}$ from atomic coordinates are continuous and differentiable, the derivatives of the feature vector components with respect to atomic positions can also be computed. The feature vectors $\mathbf{x}$ are then used as inputs to a GP model to fit the global molecular property such as the potential energy of a molecular configuration. On the other hand, the derivatives of the power spectra with respect to atomic positions are used as inputs to a GP model to fit local properties such as atomic forces. Since recently, a variety of efficient implementations for computing the invariants emerged, such as \textsc{Librascal} \cite{willatt2019atom}, \textsc{Quip} \cite{Bartok2010-pw}, and \textsc{Dscribe} \cite{dscribe}. While we briefly described the simplest invariant representation that captures the $3$-body correlations, a more expressive representations capturing higher-order correlations sometimes are required, we refer the reader to \cite{willatt2019atom} for a detailed review.

\subsection{Scalable Variational Gaussian processes}

Gaussian processes offer rich non-parametric models for regression and classification tasks \cite{williams2006gaussian}. A GP is fully specified by a mean function $\mu: \mathbb{R}^{d} \rightarrow \mathbb{R}$ and kernel or covariance function $k: \mathbb{R}^{d} \times \mathbb{R}^{d} \rightarrow \mathbb{R}$. Choosing a different kernel and a mean allows us to encode the prior information on the process that generated data. A prototypical example of a kernel that we will use across this paper is the scaled Radial Basis Function (RBF) kernel with Automatic Relevance Determination (ARD) of features \cite{williams2006gaussian}, given by
\begin{equation}
  k\left(\mathbf{x}-\mathbf{x}^{\prime} \mid \theta\right)= \theta_{s} \exp \left( - \frac{1}{2} \sum_{i=1}^{d} \frac{1}{\theta_{i}^{2}} \left(x_{i}-x_{i}^{\prime}\right)^{2} \right),
\end{equation}
where $\theta_s > 0$ is a scale parameter and $\theta_i > 0$ are the lengthscale parameters; this type of kernel allows to assign short lengthscales over the irrelevant dimensions in the descriptor space.

In the most common case of a Gaussian likelihood, the predictive distribution of the target value $y^{*}$ evaluated at the test location $\mathbf{x}^{*}$ can be computed in closed form
\begin{equation}
  p\left(y^{*} \mid \mathbf{x}^{*}\right)=\mathcal{N}\left(y^{*} \mid \mu_{\mathbf{f}}\left(\mathbf{x}^{*}\right), \sigma_{\mathbf{f}}\left(\mathbf{x}^{*}\right)^{2}+\sigma_{\text {obs }}^{2}\right),
\end{equation}
where $\sigma_{\mathbf{f}} (\mathbf{x}^{*})^2$ is the local latent function variance parameter and $\sigma_{\text {obs} }^2$ is a parameter representing the global observation noise. The variance of the prediction $y^{*}$ evaluated at the test point $\mathbf{x}^{*}$ is given by the sum of the global observation noise parameter $\sigma^2_{\mathrm{obs}}$ and the local latent function variance $\sigma_{\mathbf{f}} (\mathbf{x}^{*})^2$:
\begin{equation}
  \label{eq:variance}
  \sigma \left[y^{*} \mid \mathbf{x}^{*}\right] = \sigma_{\mathrm{obs}}^{2}+\sigma_{\mathbf{f}}\left(\mathbf{x}^{*}\right)^{2}.
\end{equation}

Typically, all the parameters of the GP are jointly learned with backpropagation by maximizing the marginal likelihood

\begin{equation}
  \label{eq:mll}
  p(\mathbf{y} \mid \mathbf{X})=\int p\left(\mathbf{y} \mid \mathbf{f}, \sigma_{\mathrm{obs}}^{2}\right) p(\mathbf{f} \mid \mathbf{X}) \mathrm{~d} \mathbf{f},
\end{equation}
where $\mathbf{y}$ is the vector with real-valued targets, $\mathbf{f}$ are the latent function values and $\mathbf{X}=\left\{\mathbf{x}_{i}\right\}_{i=1}^{N}$ are the $N$ input locations with $\mathbf{x}_{i} \in \mathbb{R}^{d}$. Unfortunately, expression \eqref{eq:mll} is too expensive to compute when number of training samples $N$ is becoming large, since computation of the inverse of the kernel matrix $\mathbf{K}_{N N}^{-1}$ scales as $\mathcal{O}\left(N^{3}\right)$, where $\mathbf{K}_{N N}=k(\mathbf{X}, \mathbf{X})$. This motivates the design of the approximate methods allowing for scalable GP inference.

Inducing point methods \cite{snelson2005sparse, titsias2009variational, hensman2013gaussian} offer a scalable extension to the exact GPs. The main idea behind the inducing point methods is to replace the large kernel matrix $\mathbf{K}_{N N}$ with its low rank approximation $\mathbf{K}_{M M}$ by introducing the latent variables $\mathbf{u} \in \mathbb{R}^M$ which are evaluated over the set of $M$ inducing points $\mathbf{Z} = \left\{\mathbf{z}_{m}\right\}_{m=1}^{M}$, with each $\mathbf{z}_{m} \in \mathbb{R}^d$. Inducing point locations can be set to be the learnable variational parameters of the model. In this scenario, the GP prior in \eqref{eq:mll} is augmented by

\begin{equation}
  p(\mathbf{f} \mid \mathbf{X}) \rightarrow p(\mathbf{f} \mid \mathbf{u}, \mathbf{X}, \mathbf{Z}) p(\mathbf{u} \mid \mathbf{Z}).
\end{equation}

Using Jensen's inequality we can lower bound the log joint density over the targets and inducing variables
\begin{equation}
  \begin{aligned}
    \log p(\mathbf{y}, \mathbf{u} \mid \mathbf{X}, \mathbf{Z}) & =\log \int p(\mathbf{y} \mid \mathbf{f}) p(\mathbf{f} \mid \mathbf{u}) p(\mathbf{u})  \mathrm{~d} \mathbf{f} \\
                                                               & \geq \mathbb{E}_{p(\mathbf{f} \mid \mathbf{u})}[\log p(\mathbf{y} \mid \mathbf{f})+\log p(\mathbf{u})].
  \end{aligned}
  \label{eq:jensens bound}
\end{equation}

Finally, by introducing the variational distribution $q(\mathbf{u})$ over the inducing variables, we come up with the following lower bounds for the evidence $p(\mathbf{y})$ \cite{hensman2015scalable, hensman2013gaussian}
\begin{equation}
  \begin{aligned}
    \log p(\mathbf{y}) & \geq \mathbb{E}_{q(\mathbf{u})}[\log p(\mathbf{y}, \mathbf{u} \mid \mathbf{X}, \mathbf{Z})]+H[q(\mathbf{u})]                                                            \\
                       & \geq \mathbb{E}_{q(\mathbf{u})}\left[\mathbb{E}_{p(\mathbf{f} \mid \mathbf{u})}[\log p(\mathbf{y} \mid \mathbf{f})]\right]-\mathrm{KL}[q(\mathbf{u}) \| p(\mathbf{u})],
  \end{aligned}
  \label{eq:evidence elbo}
\end{equation}
where $H[q(\mathbf{u})]$ is the entropy term and $\mathrm{KL}[q(\mathbf{u}) \| p(\mathbf{u})]$ is the Kullback-Leibler divergence between the distributions $q(\mathbf{u})$ and $p(\mathbf{u})$.

In the most common scenario, the variational distribution over the inducing variables $\mathbf{u}$ is given by the multivariate normal $q(\mathbf{u})=\mathcal{N}(\mathbf{m}, \mathbf{S})$ parametrized by mean vector $\mathbf{m} \in \mathrm{R}^M$ and covariance matrix $\mathbf{S} \in \mathrm{R}^{M \times M}$. Depending on the desired expressivity of the GP model and complexity of the data, we can choose to learn either the full positive semidefinite symmetric covariance matrix with non-zero non-diagonal elements, use the mean-field approximation by restricting $\mathbf{S}$ to be diagonal, or even collapse the multivariate normal $q(\mathbf{u})$ to a Dirac delta distribution by taking the formal limit $\mathbf{S} \rightarrow 0$.

Next, we briefly present several ELBO-based approximations to the high-cost marginal log-likelihood objective that enable the scalable GP inference. Note how all three objectives will be naturally written as the sum of the data-fit term and the regularization term, thus explicitly targeting the model close to the desired prior $p(\mathbf{u})$ that fits the observed data. It is crucially important that all three objectives allow for data subsampling, therefore, enabling us to use the stochastic gradient-based optimization methods \cite{hoffman2013stochastic}. The first approximation directly extends \eqref{eq:evidence elbo} by adding a regularization parameter $\beta$, we refer to it as variational ELBO objective:
\begin{equation}
  \label{eq:elbo}
  \begin{aligned}
    \mathcal{L}_{\mathrm{ELBO}} = \sum_{i=1}^{N} \mathbb{E}_{q\left(f_{i}\right)}\left[\log p\left(y_{i} \mid f_{i}\right)\right] - \beta \mathrm{KL}[q(\mathbf{u}) \| p(\mathbf{u})],
  \end{aligned}
\end{equation}
where $q\left(f_{i}\right) := \int p\left(f_{i} \mid \mathbf{u}, \mathbf{x}_{i}\right) q(\mathbf{u}) \mathrm{~d} \mathbf{u}$ and $\beta$ is a regularization parameter.


Another objective based on ELBO, referred to as $\gamma$-robust \cite{knoblauch2019robust}, aims at providing robustness to the training of GPs with respect to model misspecification and parameters learning, where the log-likelihood term in ELBO is replaced with $\gamma$-divergence:
\begin{equation}
  \label{eq:gamma}
  \begin{aligned} 
    \mathcal{L}_{\gamma} =  \sum_{i=1}^{N} \mathbb{E}_{q(f_i)}\left[-\frac{\gamma}{\gamma-1} \frac{p\left(y_{i} \mid f_i \right)^{\gamma-1}}{\int p\left(y \mid f_i \right)^{\gamma} \mathrm{~d} y}\right]  -\beta \mathrm{KL}[q(\mathbf{u}) \| p(\mathbf{u})],
  \end{aligned}
\end{equation}
where $\gamma$ is a hyperparameter and is set to be equal to $1.03$ for the rest of the paper.

The third objective, proposed by \cite{jankowiak2020parametric}, aims at improving the uncertainty estimates of the latent function and is referred to as predictive log-likelihood objective:
\begin{equation}
  \label{eq:pll}
  \mathcal{L}_{\mathrm{PLL}} = \sum_{i=1}^{N} \log \mathbb{E}_{q(f_i)}\left[ p\left(y_{i} \mid f_{i}\right) \right] -\beta \mathrm{KL}[q(\mathbf{u}) \| p(\mathbf{u})].
\end{equation}
For the presented objectives that approximate the evidence, we can calculate the derivatives efficiently and perform the parameter learning in a black-box manner \cite{ranganath2014black}.

In all of the above approximations of ELBO, the first term, which ensures fitness to the training data, is tightly balanced by the second regularization term, which additionally biases the search towards variational distributions with higher entropy, as can be seen from the first bound in \eqref{eq:evidence elbo}. The KL divergence term may also be replaced by the more general R\'enyi divergence \cite{knoblauch2019generalized}, thus allowing for wider classes of variational distributions to be explored.

\section{Experiments}

In this section, we provide numerical experiments aiming to show the benefits of using variational GPs for fitting potential energy and atomic forces for two molecular systems. The first system is given by the artificial example of isomers of a \num{12}-atom molecule, illustrated in Fig. \ref{fig:conformers}. For this system we evaluate the performance of the variational GP models with respect to the choice of the training objective, approximations of the variational distributions, and number of inducing points. The second system is given by the large molecular crystal \cite{doi:10.1021/jacs.3c08909}, for which we show that variational GP models can efficiently learn a sparse representation of the derivatives of the SOAP feature vectors and provide predictions of atomic forces on par with exact GP models.

\subsection{Fitting energies of molecular isomers}
\label{sec:fitting energies of molecular isomers}

\begin{figure}[!t]
  \centering
  \includegraphics[width=0.5\linewidth]{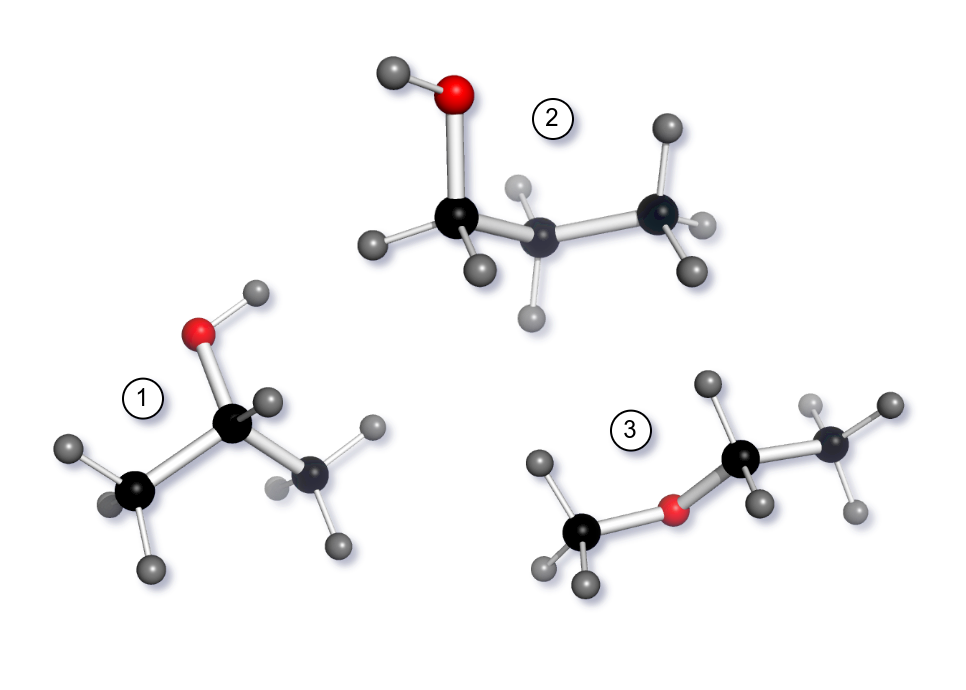}
  \caption{Structural isomers of \ce{C3H8O}.}
  \label{fig:conformers}
\end{figure}

Here we show that using the SVGP model augmented by the variationally learned inducing points can improve the prediction quality, especially on the configurations that have not been represented by the initialization selection of data points from the training set. We proceed with the example of fitting the potential energies of the conformations from the MD trajectories of three structural isomer molecules, as illustrated in Fig. \ref{fig:conformers}, with a collection of variational GPs. The target values $\mathbf{y}$ that we aim to fit in this example are the normalized potential energies of molecules from Fig. \ref{fig:conformers}. For each conformer, we generated the short \num{20} ps MD trajectory, which was then divided into the contiguous train and test parts. Each molecular conformation was mapped to the $720$-dimensional feature vector in molecular descriptor space, i.e. the input dimension for the model is $d=720$. The details of the preparation of the training dataset and details of the training are given in Appendix \ref{app: fitting energies of molecular isomers}. Here we present our main findings.

To compare different variational GP models and training strategies described in the previous section, we built a collection of $3 \times 3 \times 3 \times 2 = 54$ models corresponding to: three objective functions given by the ELBO approximations by Eqs. \eqref{eq:elbo}, \eqref{eq:gamma} and \eqref{eq:pll}; three types of normal variational distributions: one with learning the full covariance matrix, one with diagonal covariance (mean-field approximation), and one with delta covariance, where only the mean vector of the normal distribution is learned; three numbers of inducing points of \num{100}, \num{500} and \num{1000} points; two regimes defining whether the inducing point vector components are learnable parameters or not. Each of the \num{54} models was trained \num{3} times to ensure the stability of convergence. Empirically we found little dependence on the learning quality and uncertainty estimates of varying the regularization parameter $\beta$ within the approximations of ELBO and we used $\beta=0.1$ in all the experiments. 

First, in Table \ref{table:inducing points}, we show that SVGP framework augmented by variational inducing points $\mathbf{Z}$ can successfully learn to represent different types of molecules. We initialized the inducing points by random selection of the training data points corresponding only to the first molecule and then trained the SVGP model using all the training data points including also second and third molecules. In Table \ref{table:inducing points} we compare two types of training strategies: the first one is given by the fixed inducing points (FIP) where the inducing points are initialized by the configurations of the first molecule and are not learned during the training, and the second strategy is given by the variational inducing points (VIP) where the inducing points are learned during the training. In Table \ref{table:inducing points} we provide the best root mean square error (RMSE) score among all the training runs for each number of inducing points. Clearly, the variational learning of the inducing points significantly improved the predictions for all three molecules. The gain in the prediction performance by learning of the inducing points is greater for the low number of inducing points, which could be simply because too few initial inducing configurations failed to represent a large part of the training dataset; while introducing an additional $100 \times 720 = 72,000$ parameters within the model makes it more expressive. Although the inducing points were initialized by the configurations of the first molecule, the learned high-dimensional inducing point vectors could represent well configurations of the second and third molecules.  

\begin{table}[!htp]
  \centering
  \begin{tabular}{rrrr}
         & \multicolumn{3}{c}{FIP/\textbf{VIP}} \\
    IP \ \ \ & $RMSE_1$      & $RMSE_2$ & $RMSE_3$   \\  \cline{1-4} 
    100 \ \ \ & 0.049/\textbf{0.027}   & 0.057/\textbf{0.028}   & 0.100/\textbf{0.044}    \\
    500 \ \ \ & 0.032/\textbf{0.018}   & 0.046/\textbf{0.018}   &  0.079/\textbf{0.038}    \\
   1000 \ \ \ & 0.025/\textbf{0.018}  & 0.039/\textbf{0.018}   & 0.081/\textbf{0.037}              
  \end{tabular}
  \caption{\label{table:inducing points} Best root mean square error (RMSE) score among training runs for test sets from different conformers   (lower is better). IP stands for the number of Inducing Points, FIP and VIP stand for the Fixed and Variational Inducing Points, respectively. Inducing points were initialized by random sampling from the configurations of the first molecule. Units are the normalized potential energies.}
\end{table}

\begin{figure}[!t]
  \centering
  \subfloat[$R_{\Sigma}$]{\includegraphics[width=0.48\textwidth]{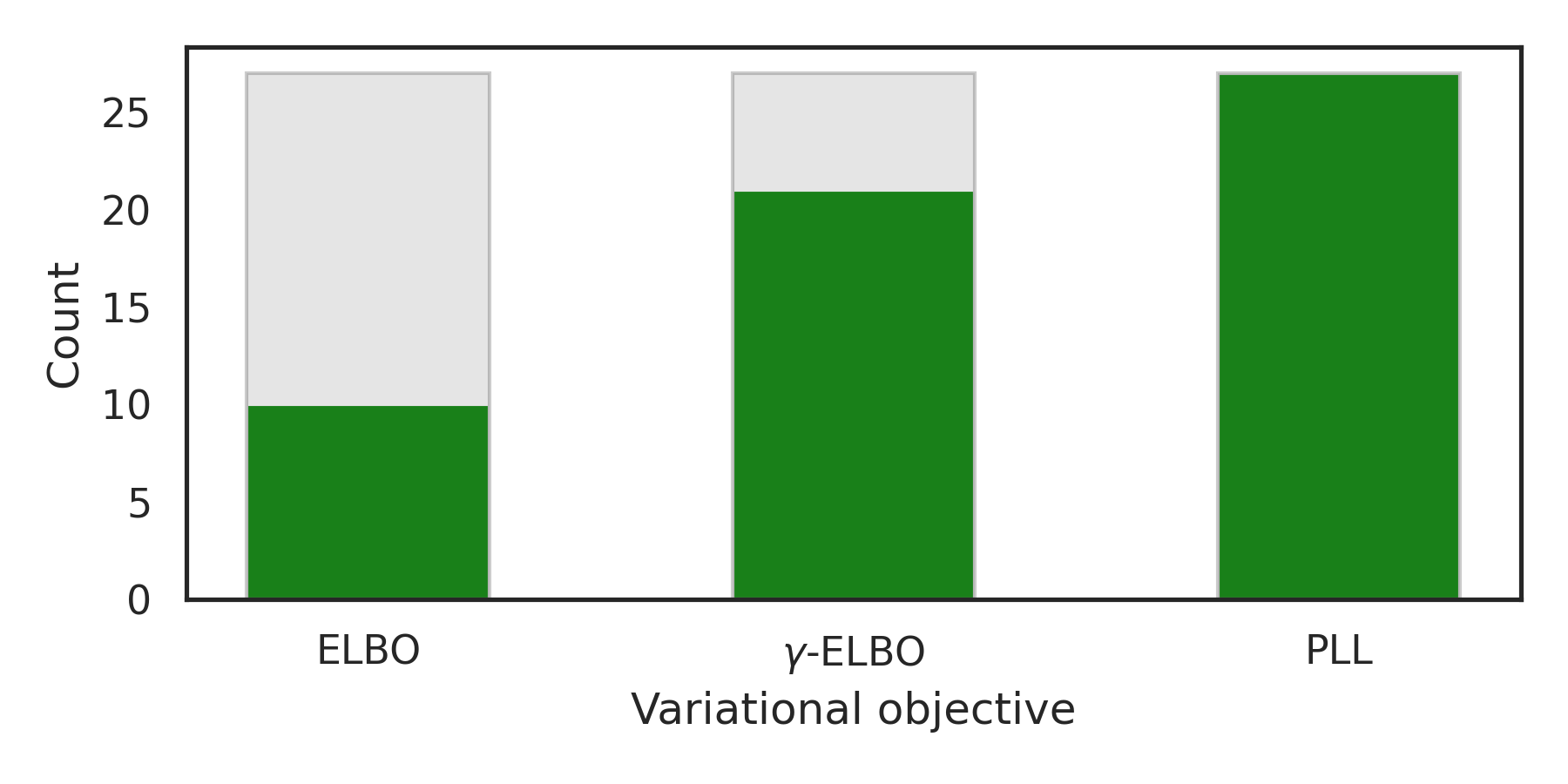}}
  \hfill\null
  \subfloat[$RMSE_{\Sigma}$]{\includegraphics[width=0.48\textwidth]{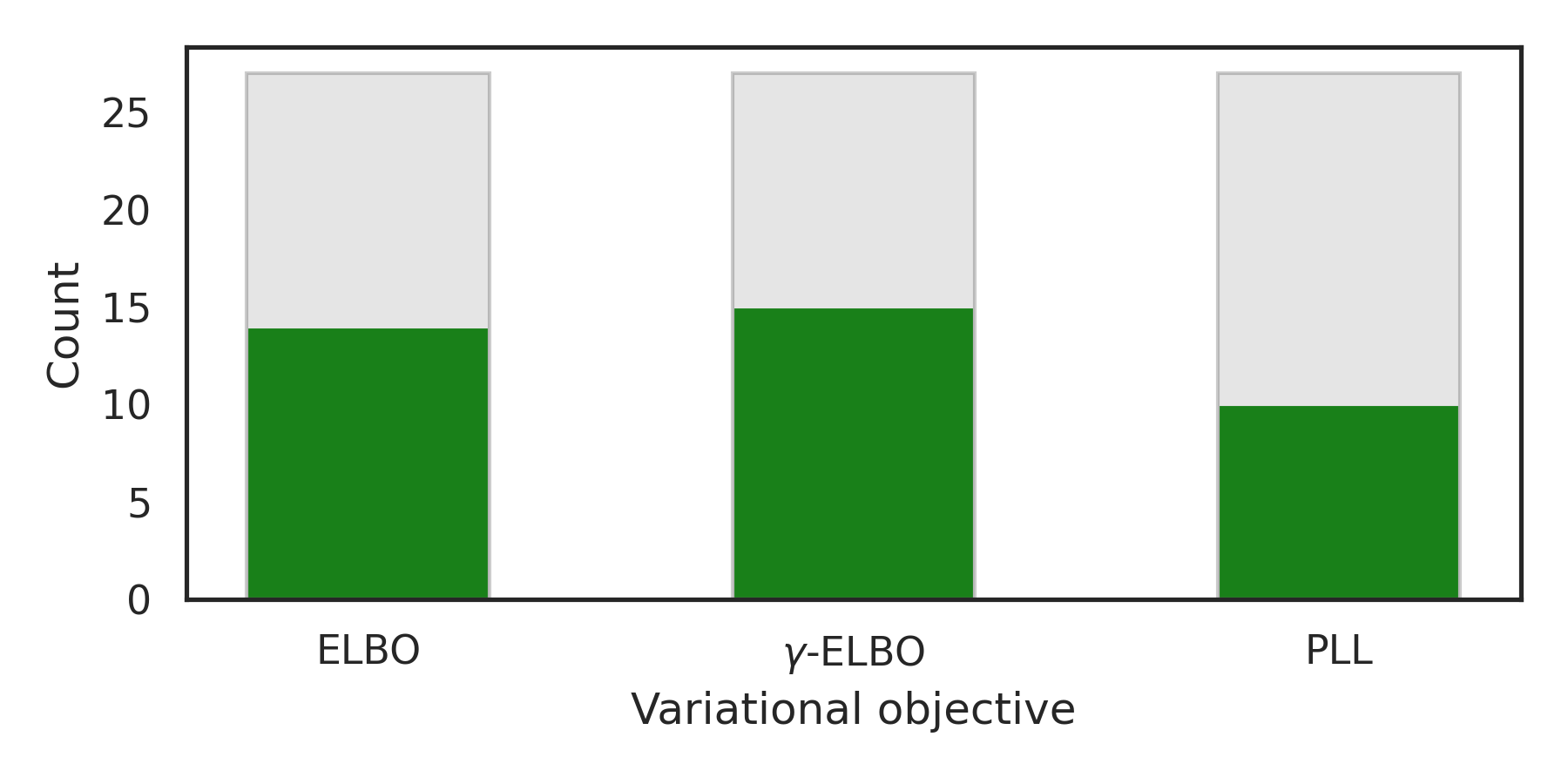}}
  \hfill\null
  \subfloat[$R_{\Sigma}$]{\includegraphics[width=0.48\textwidth]{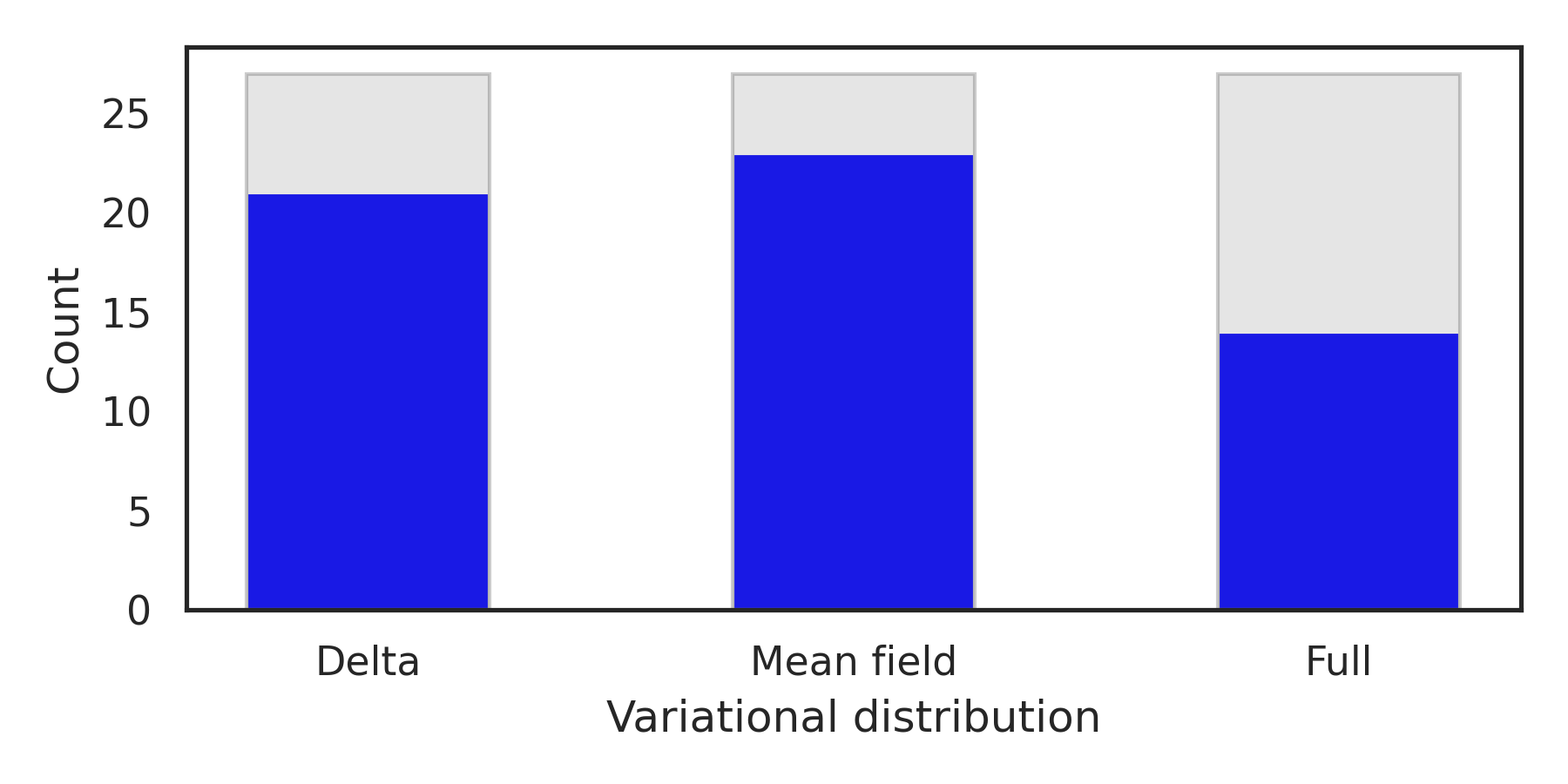}}
  \hfill\null
  \subfloat[$RMSE_{\Sigma}$]{\includegraphics[width=0.48\textwidth]{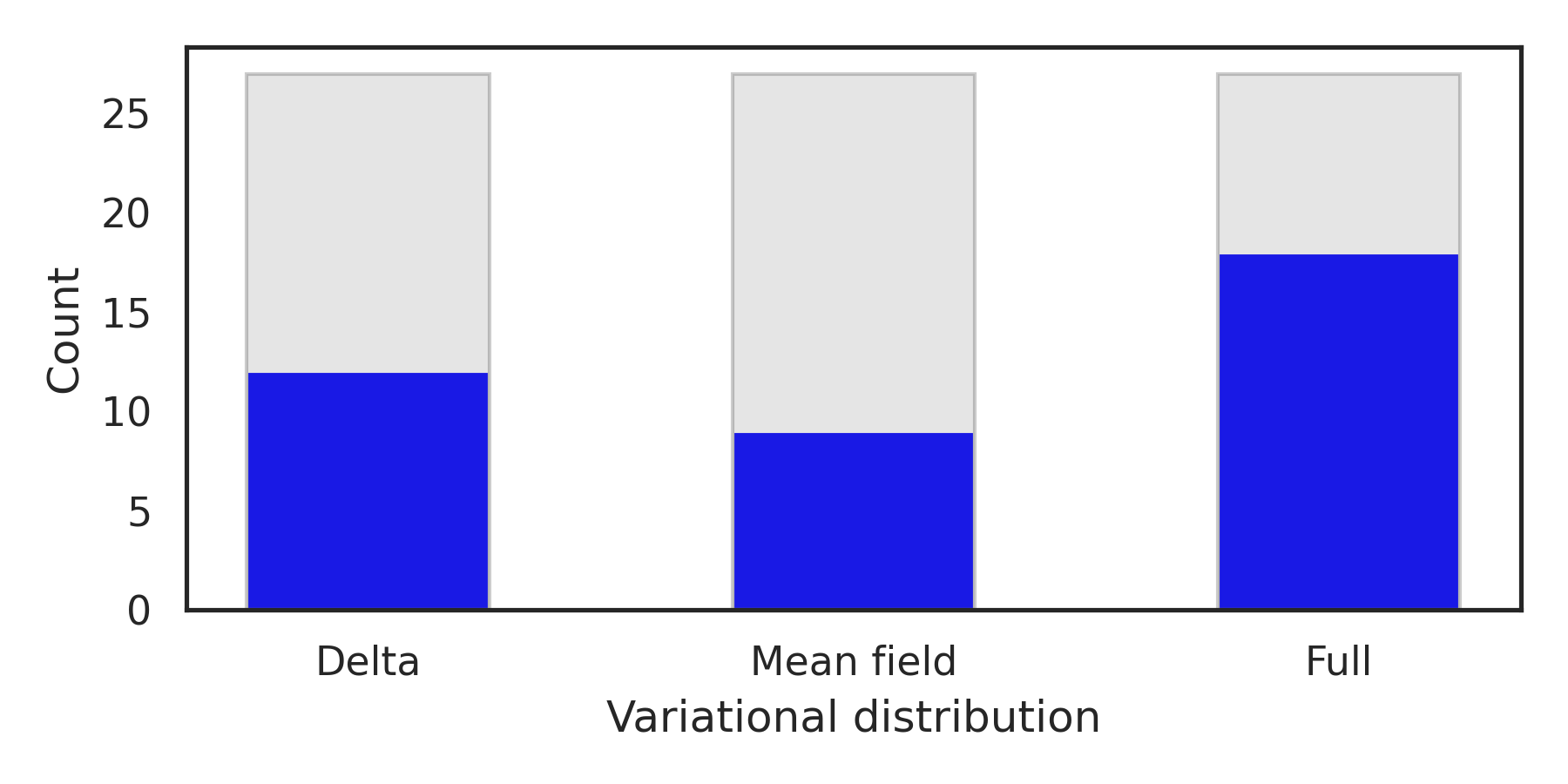}}
  \caption{}{\label{fig:elbo rmse and uncertainties} Figures (a) and (c) show the share of the training runs with best uncertainty estimates ($R_{\Sigma}<\epsilon_R$) for all three conformers, while figures (b) and (d) show the share of the training runs with best RMSE scores ($RMSE_{\Sigma}<\epsilon_{RMSE}$) for all three conformers.}
\end{figure}

To check how well GP models can distinguish uncertainties on the train and test datasets, we introduce a simple metric measuring the ratio of the mean variance of the training subset of configurations $T$ to the mean variance of the test subset of configurations $S$:
\begin{equation}
  R(T,S) = \frac{ |S| \sum_{i \in T } \sigma \left( \mathbf{x}^{*}_i \right)^{2} }{ |T| \sum_{j \in S} \sigma \left( \mathbf{x}^{*}_j \right)^{2} }.
  \label{eq:vars ratio}
\end{equation}
By denoting the $T_1$, $T_2$, $T_3$ and $S_1$, $S_2$, and $S_3$ the training and test subsets of configurations corresponding to the first, the second, and the third molecule, we computed the ratios $R_i = R(T_i,S_i)$. We have chosen the test sets to contain the continuations of the trajectories because, on the one hand, test configurations are somewhat similar to the subset of training configurations, therefore, making predictions on them makes sense, on the other hand, test configurations should still be different from the training configurations because they lie outside of the intervals of short MD trajectories, and, therefore, measuring uncertainty is also reasonable for them.

In Figs. \ref{fig:elbo rmse and uncertainties} (a) and (c), we provide the share of training runs with corresponding training objectives and variational distributions that satisfied the criterion $R_{\Sigma} = \sum_{i=1}^3 R(T_i,S_i) < \epsilon_R $, with $\epsilon_R = 0.87 \cdot 3 = 2.61$, corresponding approximately to the top $30\%$ of the score among all runs. In simple terms, we filtered training runs with the best uncertainty ratios for all three conformers. In Figs. \ref{fig:elbo rmse and uncertainties} (b) and (d), we provide the share of training runs with best RMSE scores, i.e. runs for which $RMSE_{\Sigma}= \sum_{i=1}^3 RMSE_i < \epsilon_{RMSE} $, with $\epsilon_{RMSE} = 0.028 \cdot 3 = 0.084$, also corresponding approximately to the top $30\%$. The histograms of the $RMSE_{\Sigma}$ and $R_{\Sigma}$ scores for variational objectives and variational distributions for all the training runs are provided in Appendix, Fig. \ref{fig:histogram rmse and ratios}. While the predictive log-likelihood objective provides the best uncertainty separation, it comes at the slight expense of the predictive quality. Numerical experiments additionally showed that mean-field variational distribution provides the best uncertainty separation, while the full covariance variational distribution provides the best predictive quality.

The improved uncertainty estimates by the predictive log-likelihood objective can be justified theoretically by the following argument, as given in \cite{jankowiak2020parametric}. For all three objectives given by eq. \eqref{eq:elbo}, \eqref{eq:gamma} and \eqref{eq:pll} the variance parameter $\sigma$ is given by \eqref{eq:variance}. The variational ELBO \eqref{eq:elbo} and $\gamma$-robust \eqref{eq:gamma} objectives tend to underestimate the latent function variance by attributing most of the uncertainty to the global observation noise $\sigma^2_{\mathrm{obs}}$. This is because the data fit term in \eqref{eq:elbo} contains the term proportional to $\frac{1}{\sigma_{\mathrm{obs}}^{2}}\left|y_{i}-\mu_{\mathbf{f}}\left(\mathbf{x}_{i}\right)\right|^{2}$, while in the predictive log-likelihood objective the data fit term contains the term proportional to $\frac{1}{\sigma_{\mathrm{obs}}^{2}+\sigma_{\mathbf{f}}\left(\mathbf{x}_{i}\right)^{2}}\left|y_{i}-\mu_{\mathbf{f}}\left(\mathbf{x}_{i}\right)\right|^{2}$, thus equilibrating the scales of the learned global and local contributions to the variance $\sigma$.



\subsection{Fitting atomic forces of molecular crystal}
\label{sec:fitting forces of molecular crystal}

\begin{figure}[!t]
  \centering
  \includegraphics[width=0.95\linewidth]{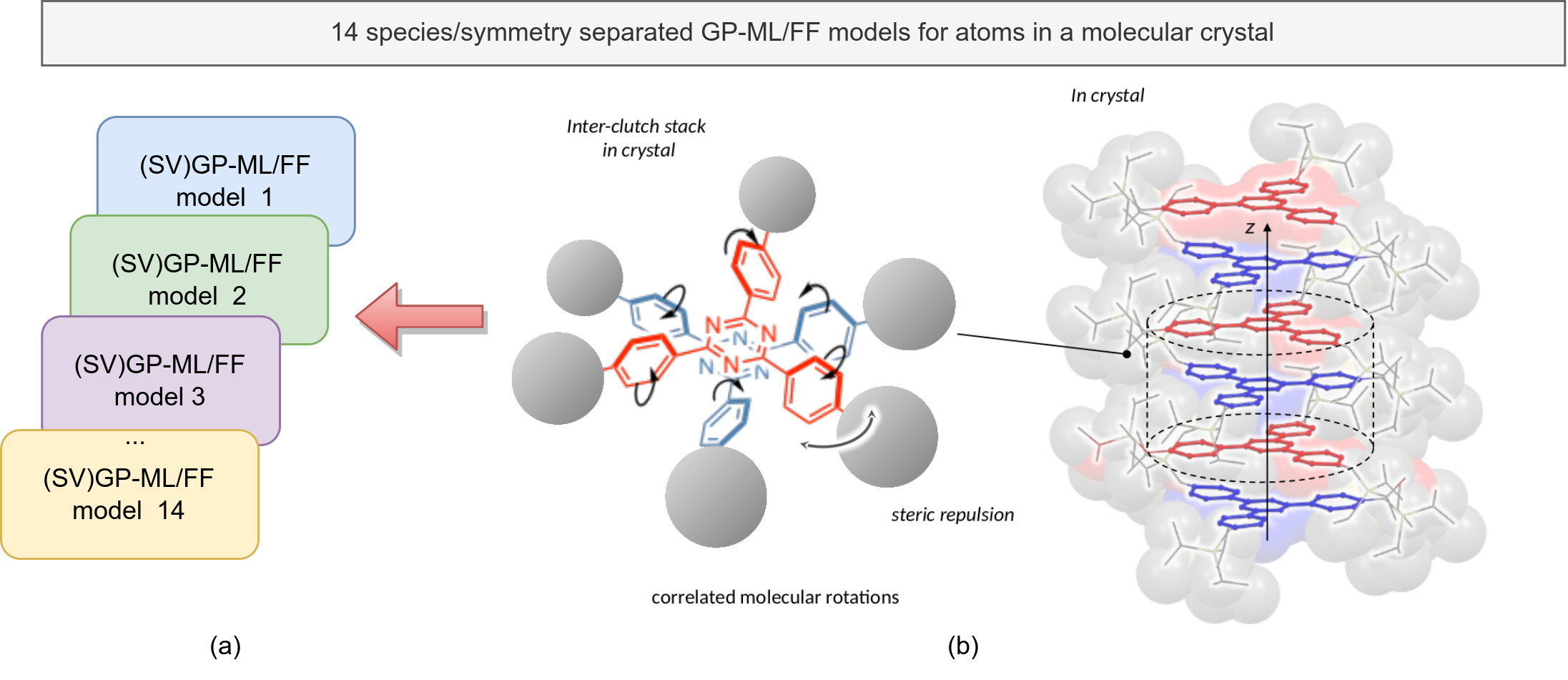}
  \caption{a) Machine learning scheme for modeling atomic forces in a molecular crystal (b). A collection of 14 GP models was trained. Each GP model correspond to atomic group with similar local atomic environments within a molecular crystal. (b) Molecular crystal system with dynamic gearing parts.}
  \label{fig:crystal diagram}
\end{figure}

\begin{figure}[!t]
  \centering
  \vspace*{0.5cm}
  \hfill\null
  \subfloat[]{\includegraphics[width=0.6\textwidth]{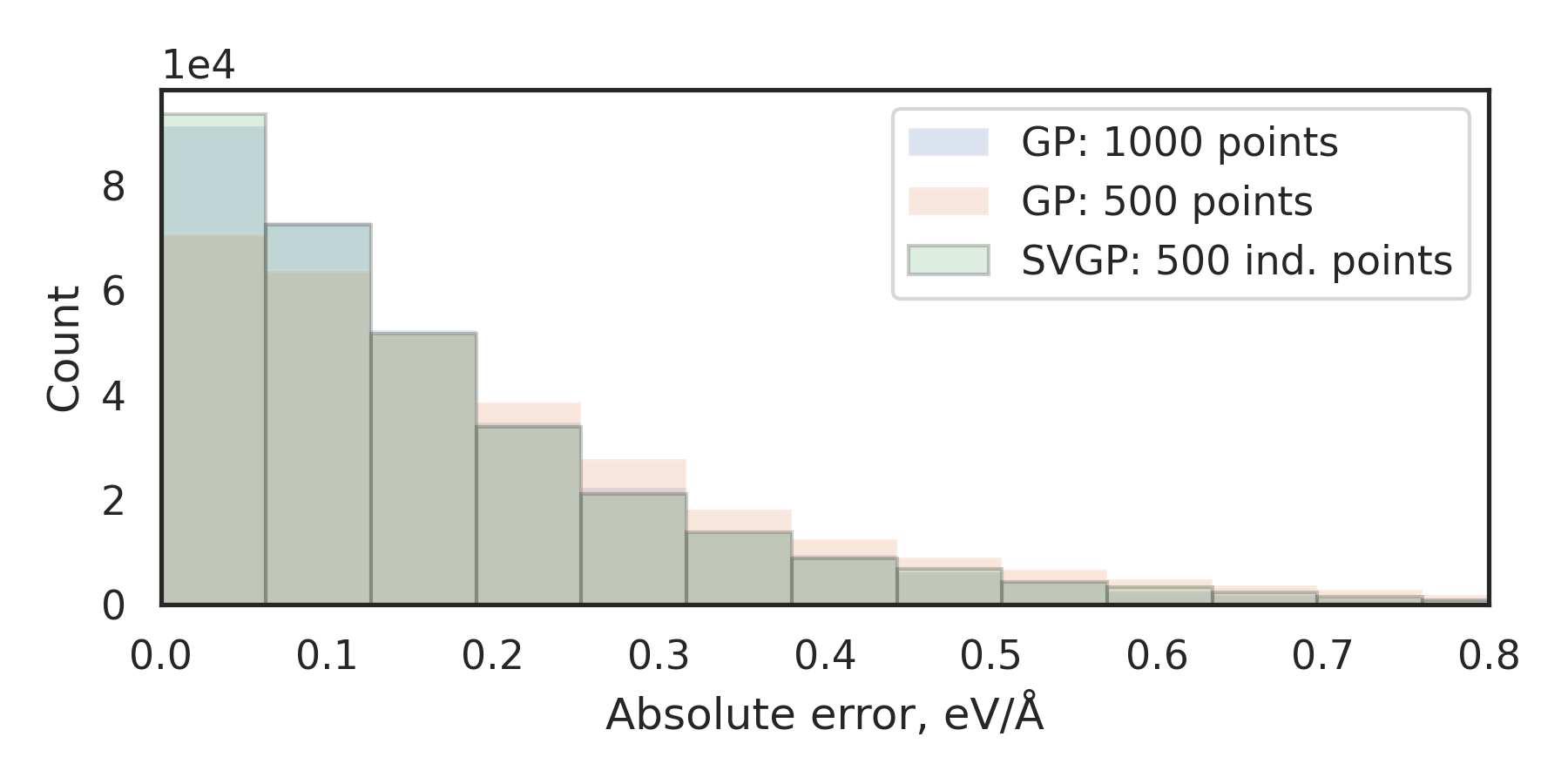}}
  \hfill\null
  \vspace*{0.5cm}
  \subfloat[]{\includegraphics[width=0.6\textwidth]{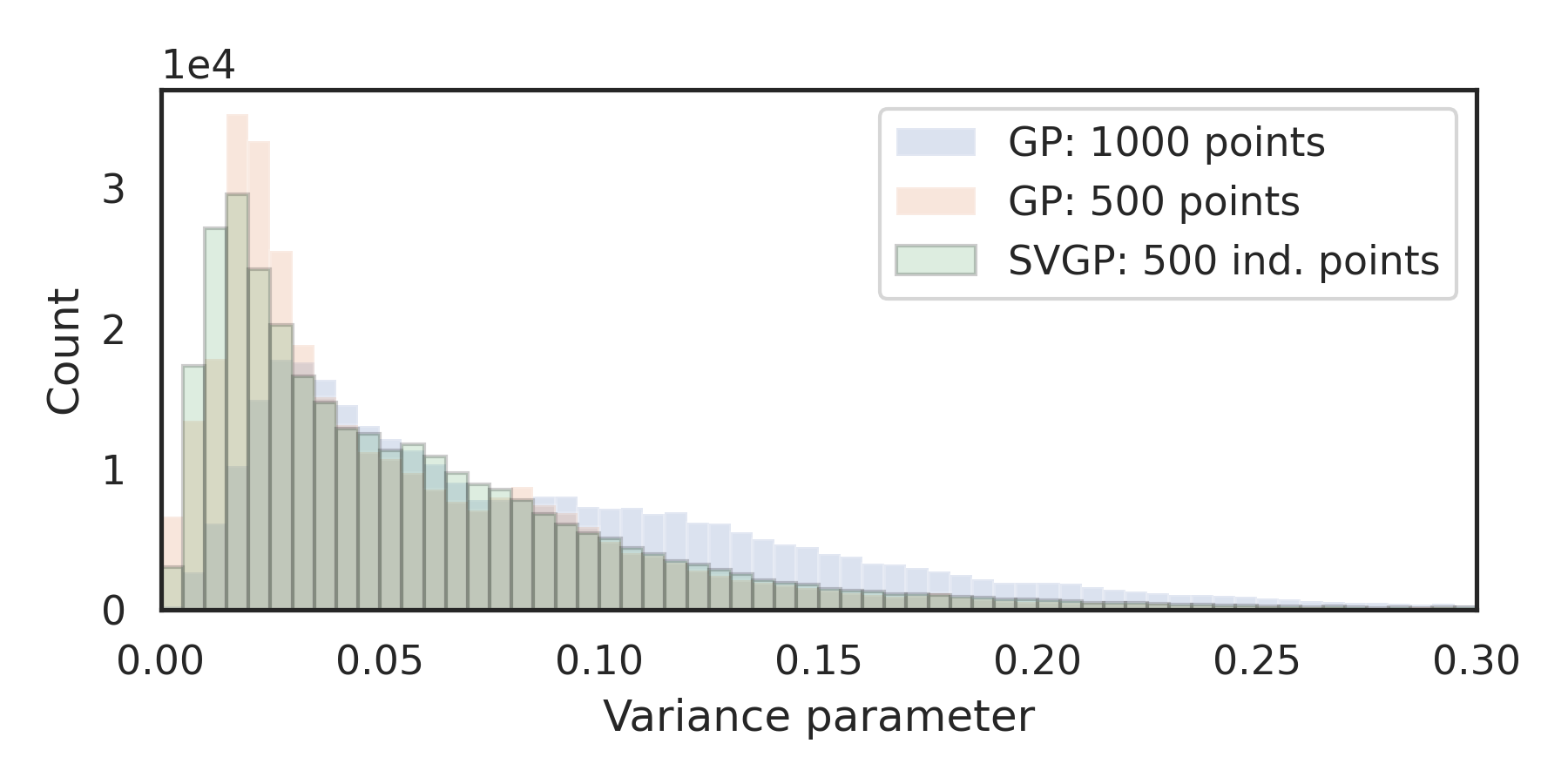}}
  \caption{}{\label{fig:forces crystal} (a) Absolute error for atomic forces on all atoms predicted by the set of SVGP models, each parametrized with 500 inducing points and trained on 1000 points, and for two exact GP models trained with 500 and 1000 points. (b) Distribution of the normalized variance parameter learned by the SVGP and exact GP models.}
\end{figure}

In this section, we provide numerical results aiming to illustrate the performance of the SVGP framework for fitting the atomic forces and compare the variational modeling framework with the exact GPs. Based on the performance results from the example of molecular conformers, here we chose to showcase the performance of the predictive log-likelihood objective for fitting the atomic forces of a much larger molecular system. Next we provide the numerical results for fitting the atomic forces of a large crystalline system with dynamic gearing parts \cite{doi:10.1021/jacs.3c08909}. 

Based on the symmetries of local chemical environments within the molecular crystal, we split the 528 atoms of the molecular crystal within a unit cell into 14 groups and trained 14 SVGP and exact GP models as illustrated in Fig. \ref{fig:crystal diagram}. For each SVGP model, we initialized 500 inducing points by randomly sampling a total of 1000 training points for each GP model. To compare the performance of SVGP models trained with predictive log-likelihood objective \eqref{eq:pll} and exact GP trained with objective \eqref{eq:mll}, we additionally trained two sets of exact GP models: with 500 and 1000 force samples. The feature vectors that describe the local chemical environment of each atom and their derivatives with respect to atomic positions were represented by a $1200$-dimensional SOAP vectors \cite{bartok2013representing}, i.e. the input dimension for the model is $d=1200$. We used the derivatives of the SOAP vectors of an atom with respect to the position of the same atom as input feature vectors for the model. In this example, we used the full covariance matrix to represent the variational distribution. The details on the generation of the training dataset and the training procedure are given in Appendix \ref{app: fitting forces of molecular crystal}.

The kernel function used in this example was the scaled Mat\'ern kernel with Automatic Relevance Determination (ARD) of features, defined by \cite{williams2006gaussian}
\begin{equation}
    k_{\text {Matern }}\left(\mathbf{x}_1, \mathbf{x}_2\right) = \theta_{s} \frac{2^{1-\nu}}{\Gamma(\nu)}(\sqrt{2 \nu} d')^\nu K_\nu(\sqrt{2 \nu} d')
\end{equation}
where $\Gamma$ is a Gamma function, $K_\nu$ modified Bessel function of the second kind, $d' =  \frac{1}{2} \sum_{i=1}^{d} \frac{1}{\theta_{i}^{2}} \left(x_{i}-x_{i}^{\prime}\right)^{2}$ is a weighted distance between points, $\theta_s > 0$ is a scale parameter, $\theta_i > 0$ are the lengthscale parameters, and in this example we used $\nu=2.5$.

To train the SVGP and exact GP models, we generated the training dataset by performing \textit{ab initio} molecular dynamics simulation (simulation and training details are given in Appendix \ref{app: fitting forces of molecular crystal}). To evaluate the performance of the SVGP and exact GP models, as a test trajectory we took the trajectory and corresponding atomic forces of correlated barrier crossing by all the 12 gears within a unit cell as illustrated in Fig. \ref{fig:crystal diagram} (b). This trajectory was used in the paper \cite{doi:10.1021/jacs.3c08909} for computational estimates of the rotational barriers of gears.

\begin{table}[!tp]
  \centering
  \begin{tabular}{lcccc}
    Model                   & \ \ RMSE \ \ & Outliers   & \\ \cline{1-3}
    GP: 500 points          & 0.07822            & 4858                  & \\
    GP: 1000 points         & 0.05045            & 1977                  & \\
    SVGP: 500 ind. points   & \textbf{0.05008}   & \textbf{1518}                  &
  \end{tabular}
  \caption{\label{table:performance on forces} Performance of GP models for atomic force predictions of a molecular crystal trained with variational and exact GPs. Outliers refer to the number of test predictions with an absolute error greater than $0.8$ \si{eV}/\si{\angstrom}.}
\end{table}

In Fig. \ref{fig:forces crystal} we provide numerical results for three modeling settings: (1) collection of 14 SVGP models, each with 500 inducing points and trained on 1000 points; (2) collection of 14 exact GP models trained on 1000 randomly sampled points; (3) collection of 14 exact GP models trained on the 500 points. The inducing points of SVGP models were initialized by randomly sampling 500 points out of 1000 training data points. The SVGP and exact GP models were trained similarly with backpropagation by optimizing the objective functions \eqref{eq:mll} and \eqref{eq:pll} respectively. 

In Fig. \ref{fig:forces crystal} (a) we provide the absolute errors of the predicted atomic forces for three modeling setups. Even though SVGP used 500 variationally learned inducing points, the absolute error for the test dataset for the SVGP models is on par with the exact GPs trained on 1000 points. This could be related to the fact that each SVGP model has additional $500 \times 1200 = 600,000$ variationally learned parameters allowing the SVGP models to find a more optimal representation of the dataset. In Fig. \ref{fig:forces crystal} (b) we provide the histogram of the learned GP variance parameter (normalized to be in $[0,1]$ interval) for all atoms within a unit cell. Distributions of the variance parameters differ significantly for the SVGP and exact GP models, with exact GP models trained on 1000 points tending to have relatively higher normalized variance than SVGP models. Additionally, in Table \ref{table:performance on forces} we provide the RMSE scores for the atomic force predictions and the number of outliers for the SVGP and exact GP models. The SVGP models with 500 inducing points and trained on 1000 points gave a similar RMSE score to the exact GP models trained on 1000 points, however, with fewer outlier predictions.


\subsection{Initialization of inducing points and stochasticity of the training}
\label{sec:initialization of inducing points}

Although the training of the variational GP model has been cast in a black-box manner, we still need to comment on possible pitfalls. One possible problem while training SVGPs is the initialization of the inducing points $ \mathbf{Z}$. For example, we observed that optimization of the components of high-dimensional inducing point vectors, initialized by random draws from the normal distribution $\mathcal{N}(0,1)$, is typically stuck in some local minimum and the trained model performs poorly. Therefore proper initialization of inducing point vectors is needed. In this work, we initialized the inducing points by randomly sampling from the training data points. A better initialization strategy could be to sample the farthest points from the training set \cite{moenning2003fast} or use the $k$-means clustering algorithm. Even better methods for choosing samples could also take into account the intrinsic topology of the dataset \cite{dufresne2019sampling} since descriptors of conformers of molecular dynamics trajectory in principle follow a continuous path in a high-dimensional descriptor space. Additionally, since training datasets are frequently obtained by the MD simulations, the configurations are often sampled non-uniformly. To mitigate this issue, alternative approaches have been proposed for generating the training dataset \cite{waters2021energy,waters2022benchmarking}.

Another issue is whether inducing points can be learned by gradient descent, and related to what kind of stochastic gradient descent algorithm is used for iterations. In this paper, for both examples, we used full-batch training with Adam optimizer \cite{kingma2014adam} and multi-step gradient optimization with $2$ gradient step sizes. The necessity of mini-batch training for generalization is currently an open question: while some works \cite{geiping2021stochastic} claim that generalization is achievable in a full-batch regime, others \cite{ober2021promises} claim that mini-batch training is necessary for generalization.


\section{Interpolation vs. extrapolation}
\label{sec:interpolation vs. extrapolation}

It is often tacitly assumed that simple non-variational GPs make predictions on the test data points by interpolating within the training set of points, which, for the low-dimensional datasets, is mostly an obvious statement. At the same time, looking at the simulations of the MD of small molecules, like the ones from Fig. \ref{fig:conformers}, makes us think that conformational diversity during dynamics is quite restricted -- if some test snapshot is taken in between two closely sampled in time neighboring snapshots, intuitively it ``interpolates'' between its neighbors. Hence, if we want to predict some physical property, like potential energy, for a collection of test configurations, we might be lured into thinking that the GP generally will \textit{interpolate in between} the training configurations also in the descriptor space. By ``in between,'' it is typically assumed that a test point will be within the convex hull of the training points. Here we follow the arguments by \cite{balestriero2021learning}, where it was shown that for high-dimensional datasets, like the image data, learning always amounts to extrapolation. Recently, a similar kind of analysis was performed on datasets of bulk structures and crystals \cite{zeni2022exploring}, where authors additionally propose an alternative protocol to quantify the similarity of a test point to the training set of points based on the local estimated density of training points.

Next, we demonstrate, similarly to \cite{balestriero2021learning} and \cite{zeni2022exploring}, that for the MD datasets with high-dimensional descriptor representations like SOAP, the situation is quite similar, even if we form the dataset by very similar molecular configurations, the randomly sampled test configuration data point will most likely fall outside the convex hull of the training set of points. This phenomenon is due to the curse of dimensionality, it clearly shows that the meaning of interpolation and extrapolation needs to be properly adjusted to the problem at hand when working with high-dimensional data and alternative similarity measures need to be constructed.

\begin{figure}[!t]
  \centering
  \subfloat[]{\includegraphics[width=0.68\textwidth]{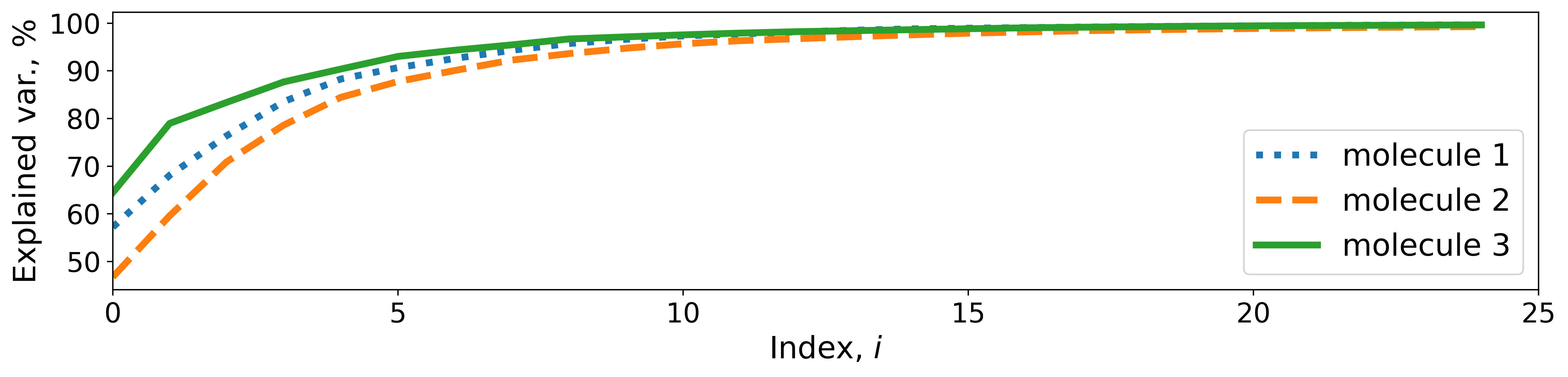}}
  \hfill\null
  \subfloat[]{\includegraphics[width=0.68\textwidth]{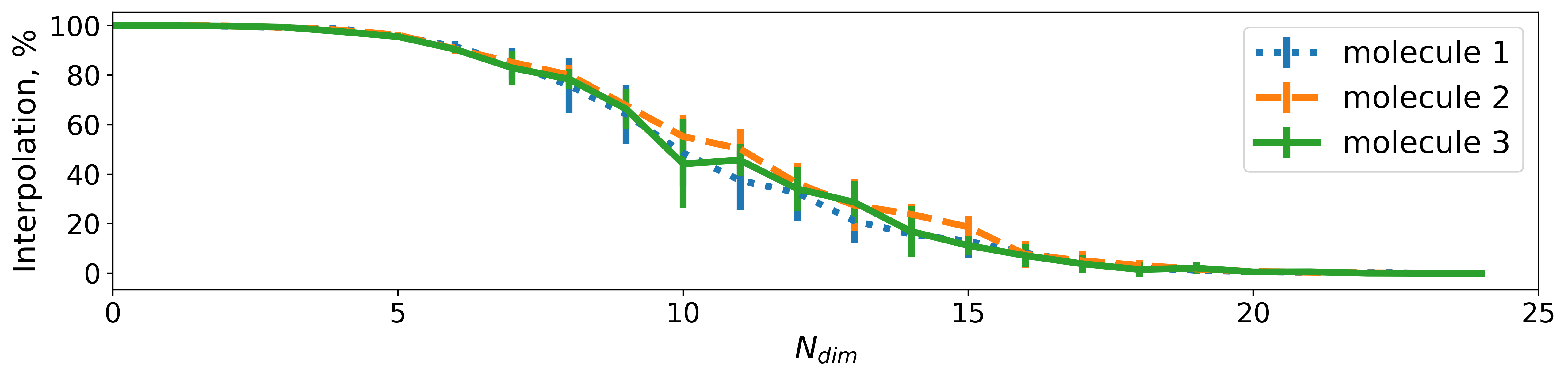}}
  \caption{}{\label{fig:singular values plot} (a) Fraction of the total variance explained by the first $i$ singular vectors. (b) Fraction of points from the test set within the convex hull of the training points for random selections of $N_{dim}$ dimensions; the standard deviations across the runs are shown with vertical bars.}
\end{figure}

We illustrate this phenomenon on the dataset of MD trajectories of the molecules from Fig. \ref{fig:conformers}. As before, for these molecules, we generated a large set of conformations simply by simulating a $100$ \si{ps} MD trajectories at $300$ K temperature using the same parameters as in the previous section. To get an estimate of the variability of the invariant vectors within the dataset, we applied the Principal Component Analysis (PCA) to the dataset containing invariant descriptor vectors for \num{10000} consecutive snapshots. In Fig. \ref{fig:singular values plot} (a), we illustrate what fraction of the total variance is explained by the first $i$ largest singular values $\sigma_j$, i.e., we plot $r_i = \sum_{j=0}^{i-1} \sigma_j^2 / \sum_{j=0}^{d-1} \sigma_j^2 $. For example, the first $15$ principal components explain $96.5 \%$, $94.7 \%$ and $97.1 \%$ of total variance for the first, the second, and the third molecule.


Next, we show what fraction of test points falls within the convex hull of training points for randomly selected subspaces in the descriptor space. First, we formed a randomized collection of test vectors by randomly selecting \num{200} configurations from the pool of \num{10000} configurations corresponding to the \num{100} \si{ps} trajectory for each molecule. Then for a collection of train and test vectors, we randomly subsampled $N_{dim}$ dimensions.

Constructing the convex hull for high-dimensional datapoints is computationally expensive, however, checking hull membership for a test point can be cast as a linear programming problem with positivity constraints. By successfully solving the linear program we can check whether a test point can be represented by a linear combination of the training points with positive weights summing up to one.

In Fig. \ref{fig:singular values plot} (b), we illustrate what fraction of test points falls within the convex hull of the training points for randomly sampled $N_{dim}$ dimensions in the descriptor space. For each number of dimensions $N_{dim}$ we made \num{10} runs with different random selections of test configurations and dimensions for random projections. To conclude, we see that due to the curse of dimensionality, in the high-dimensional molecular descriptor space, only a small fraction of predictions will be performed in the interpolation regime.

\section{Conclusion}

In this work, we performed an empirical study of applying scalable variational GPs for fitting the potential energies and atomic forces for two molecular systems. First, we demonstrated that during the training, the SVGP models can learn to represent different types of molecular configurations that were not present within the initialization set of inducing points. Then we provided a comparison of various alternatives for the SVGPs training, including different variational objectives and variational distributions. Our experiments showed that introducing and learning the latent inducing point variables in a high-dimensional molecular descriptor space with a predictive log-likelihood objective provided improved uncertainty estimates, however at a slight expense of accuracy. Finally, we illustrated on the example of a large molecular crystal system that a sparse SVGP model trained with a predictive log-likelihood objective can provide predictions of atomic forces on par with the exact GP models.

Although augmenting a GP model with a large number of variational parameters clearly showed its advantages, it also made the training process more complex. Moreover, even though the learned inducing points could represent different types of configurations, their proper initialization was necessary, and the extent to which inducing points can adapt to represent different molecular configurations is still unexplored. In future work, we aim to address the above issues as well as apply scalable variational GP models to large-scale problems.


\bibliographystyle{siam}
\bibliography{bibliography}


\clearpage
\appendix

\section{Experiment details}
\label{app:experiment details}

\subsection{Fitting energies of molecular isomers}
\label{app: fitting energies of molecular isomers}
\begin{figure}[!t]
  \centering
  \subfloat[]{\includegraphics[width=0.48\textwidth]{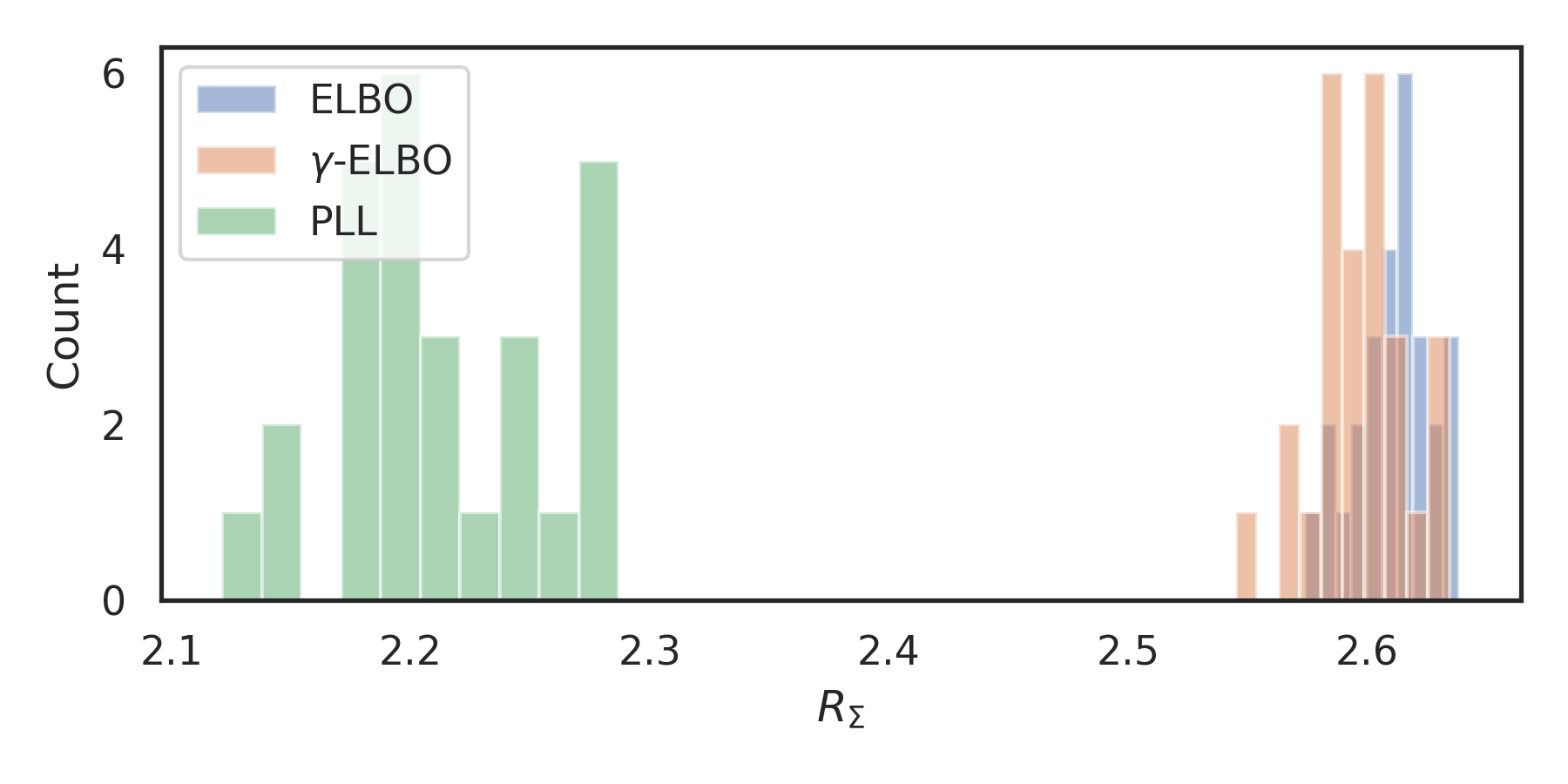}}
  \hfill\null
  \subfloat[]{\includegraphics[width=0.48\textwidth]{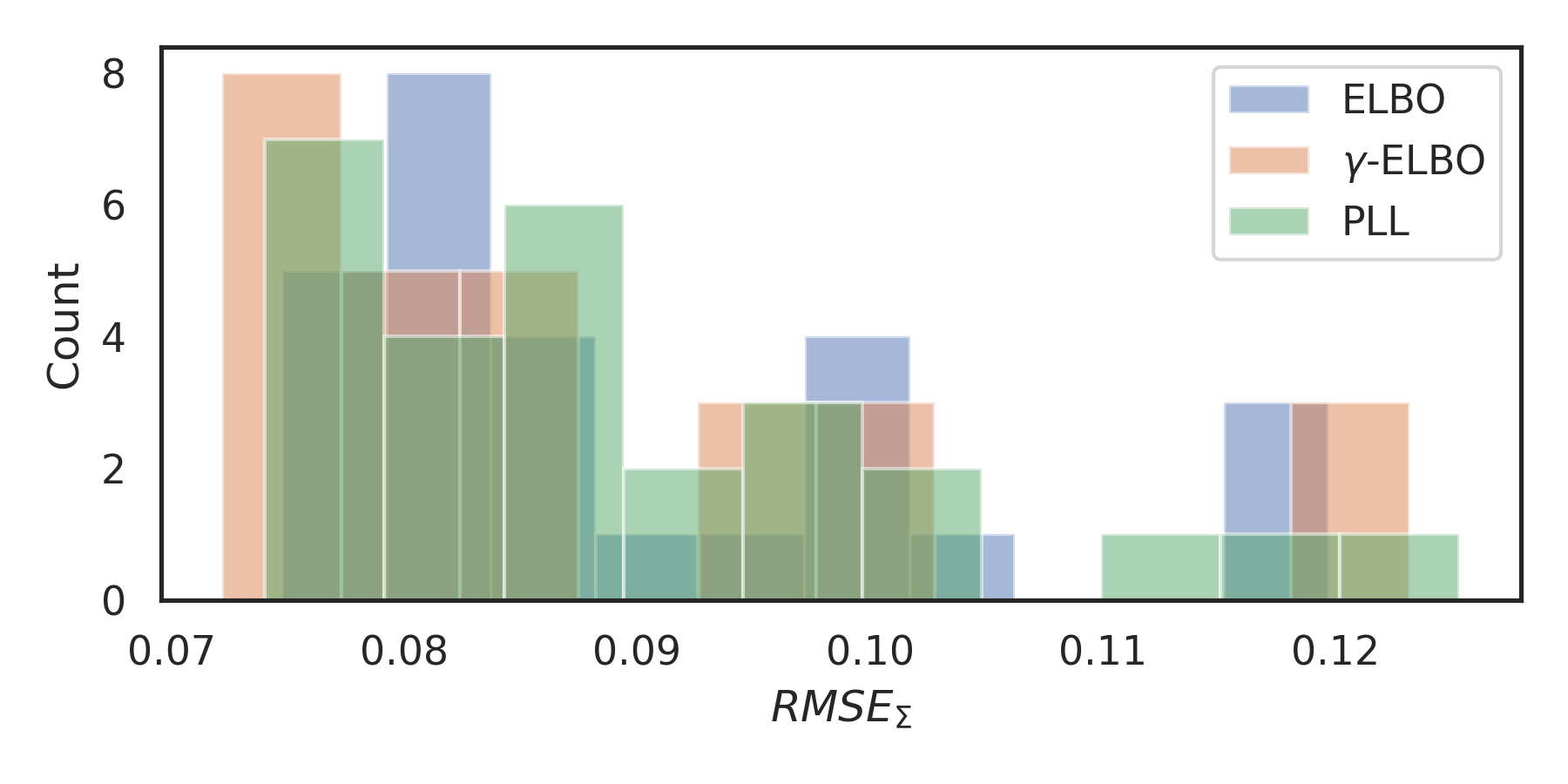}}
  \hfill\null
  \subfloat[]{\includegraphics[width=0.48\textwidth]{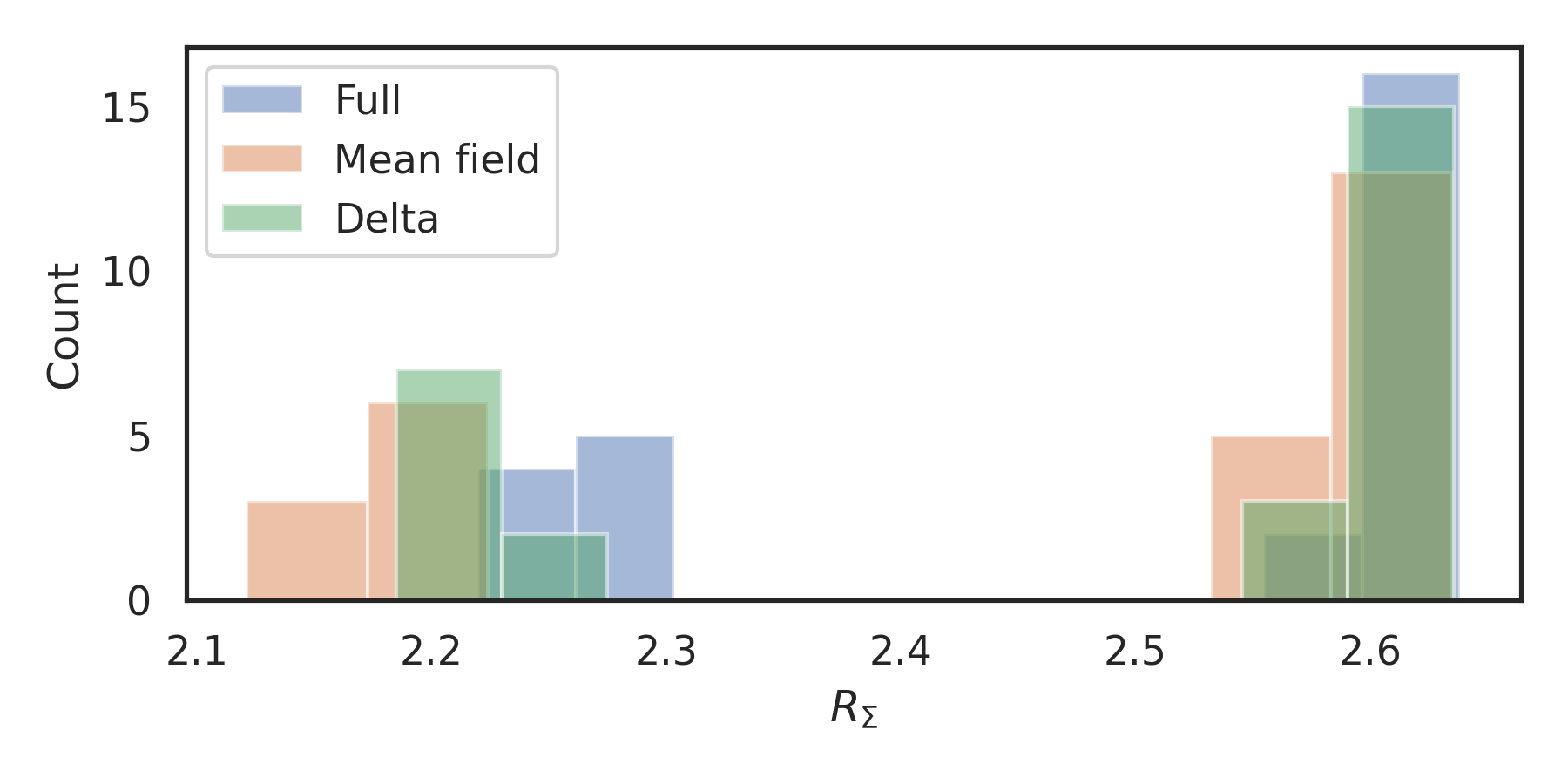}}
  \hfill\null
  \subfloat[]{\includegraphics[width=0.48\textwidth]{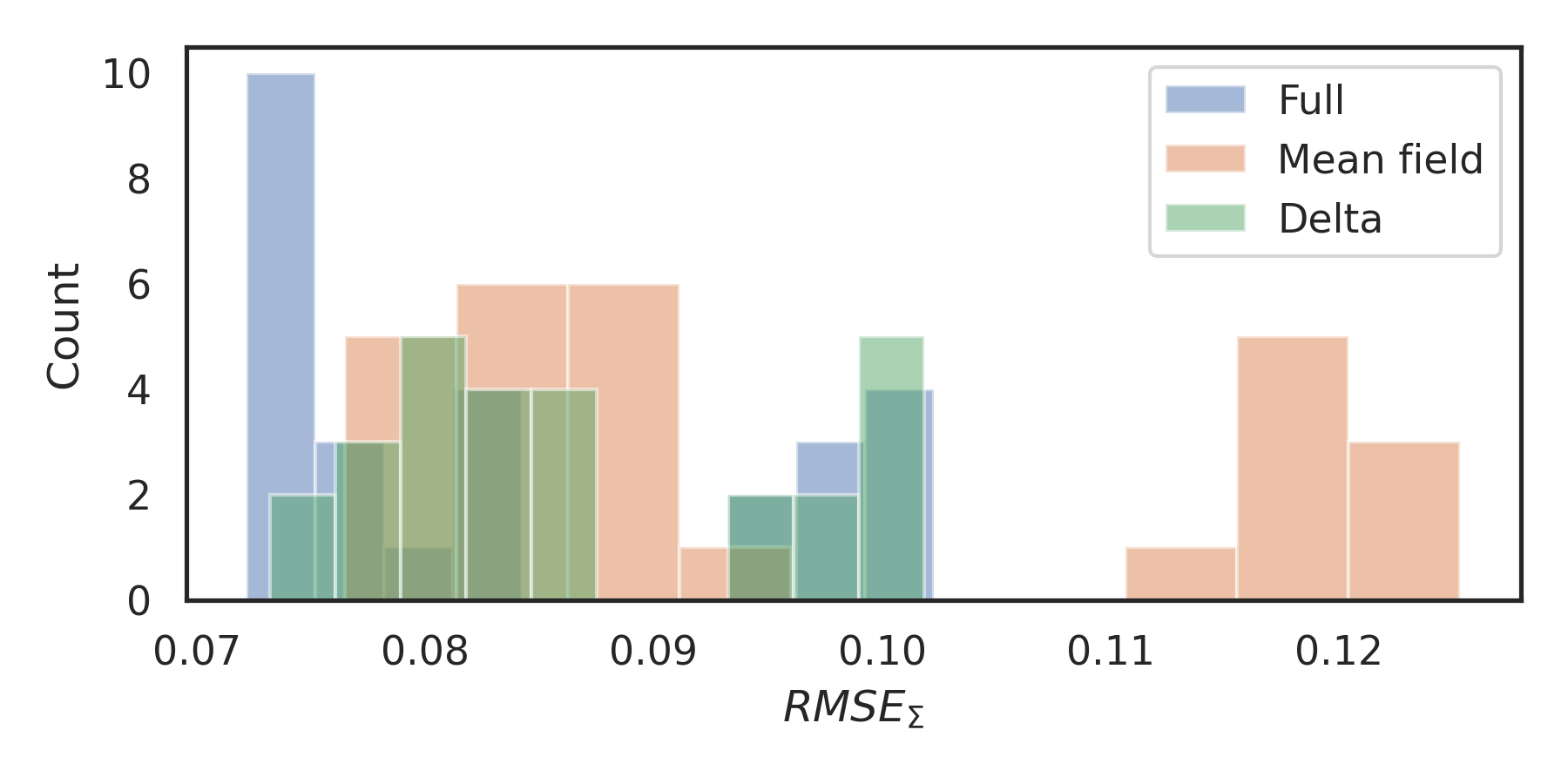}}

  \caption{}{\label{fig:histogram rmse and ratios} Figures (a) and (c) show the distribution of the ratio score $R_{\Sigma}$ for all training runs, while figures (b) and (d) provide the $RMSE_{\Sigma}$ score for all training runs.}
\end{figure}

For each molecule, we generated a short \num{20} \si{ps} MD trajectory at \num{300} \si{K} temperature with Berendsen thermostat and GFN2-xTB tight-binding Hamiltonian \cite{bannwarth2019gfn2}. Before simulation molecular geometries were optimized with the same GFN2-xTB tight-binding Hamiltonian. The equilibration period of \num{10} \si{ps} was used. All simulations were performed in vacuum. Trajectories were calculated with $0.4$ \si{fs} time step and then subsampled to $10$ \si{fs} time step \cite{mikhail_tsitsvero_2022_6496747}.  After subsampling each trajectory consisted of \num{2000} snapshots. We then selected the first \num{1600} configurations within each trajectory to be included in a training dataset, while the remaining \num{400} configurations within each trajectory, were selected for test sets. In total, the training set consists of \num{4800} configurations and there are three test sets with \num{400} configurations each. For each configuration of test/train sets, we computed the $720$-dimensional invariant SOAP vector representation \eqref{eq:outer sum} with \cite{dscribe}. The invariant vectors were computed with the following parameters: $n_{max} = 5$, $l_{max} = 5$, $r_{cut} = 5.0$ \si{\angstrom} and $\sigma = 0.5$ \si{\angstrom}. Four atoms, namely, the three carbons and the oxygen, were selected as centers for calculating the invariant vectors by Eq. \eqref{eq:outer sum}. For each configuration of MD trajectory, we extracted the potential energy given by the GFN2-xTB tight-binding Hamiltonian \cite{bannwarth2019gfn2}. The difference between the maximum and minimum energies of configurations within the training set (including all three molecules) was \num{0.8036} \si{eV}, thus we normalized the energies to fall into the interval $[ 0,1 ]$ scaling by this number. In the figures and tables within the main text, we provide the error metrics in the normalized units. Our goal was not to reach the ultimate accuracy but to illustrate the concept, for the ultimate accuracy, one may need to calculate the invariant descriptors with a greater number of radial and spherical basis functions, design the optimal shape of the cutoff function, choose optimal values of $\sigma$, as well as design more efficient ways to combine the local invariants, other than summation \eqref{eq:outer sum} -- a separate set of problems. In Fig. \ref{fig:histogram rmse and ratios} we provide the histograms of the RMSE and the variance ratios for all training runs. Tables with scores for individual isomers and used hyperparameters are provided in the supplementary material.

 The estimated training time for all \num{54} models was \num{54} hours. All variational models were trained with \num{20000} gradient steps. The starting learning rate was set to \num{0.01} for the first \num{1000} steps, then reduced to \num{0.001} until reaching the \num{10000} steps, and finally reduced to \num{0.0001} until reaching the \num{20000} steps.

\subsection{Fitting forces of molecular crystal}
\label{app: fitting forces of molecular crystal} 

To cover the conformational diversity of the molecular crystal, we calculated the \textit{ab initio} molecular dynamics trajectory with \num{5000} steps at temperature $T=2000$ \si{K} with Langevin thermostat and a time-step of \num{1} \si{fs}. The equilibration period of \num{2} \si{ps} was used. We sampled \num{1000} configurations from the trajectory and computed the SOAP vectors and their derivatives using the \textsc{Librascal} library \cite{willatt2019atom}. The parameters used to compute the SOAP vectors: $n_{max} = 4$, $l_{max} = 4$, $r_{cut} = 4.0$ \si{\angstrom} and $\sigma = 0.3$ \si{\angstrom}. The \textit{ab initio} Density-functional-theory (DFT) calculations were carried out using the Perdue-Burke-Ernzerhof (PBE) exchange-correlation functional and the projector-augmented wave (PAW) method, as implemented in the Vienna ab initio Simulation Package (VASP) \cite{kresse1996efficient, blochl1994projector, perdew1996generalized}. A plane wave basis set with an energy cutoff of \num{500} \si{eV} was used. Gaussian smearing with a width of 0.05 eV was applied to treat partial occupancies of the electronic orbitals, and an accuracy condition for energy conversion of $10^{-6}$ eV was used. Only $\Gamma$-point calculations were performed for the Brillouin zone sampling due to the large size of the supercell.

The estimated training time for all 14 models was about 2 hours. Each model was trained by $1000$ iterations with $0.1$ learning rate step, followed by tuning $1000$ iterations with a reduced learning rate of $0.01$.

\section{Computational details}
\label{app:computational details}
All the experiments were performed at NVIDIA RTX 8000 GPU. We used \textsc{GPyTorch} library for modeling scalable Gaussian processes \cite{gardner2018gpytorch} and the \textsc{DScribe} \cite{dscribe} and \textsc{Librascal} \cite{willatt2019atom} libraries for computing SOAP invariant descriptors. The ``whitening'' \cite{matthews2017scalable} transformation was used in both examples to accelerate the training of variational parameters, as implemented within \textsc{GyTorch} library. The models from both examples were trained with the Adam optimizer \cite{kingma2014adam}. The lengthscale and scale hyperparameters were initialized by standard initialization procedures within \textsc{GPyTorch} library.

\end{document}